\documentclass[manuscript,acmsmall,screen]{acmart}
\AtBeginDocument{%
  }
\usepackage{ltl}
\usepackage{todonotes}
\usepackage[algoruled,linesnumbered,vlined]{algorithm2e}
\DontPrintSemicolon
\SetKwProg{Proc}{Procedure}{}{}
\SetKw{KwTo}{:}
\usepackage{sansmath}
\usepackage{enumitem}
\usepackage{subcaption}
\usepackage{hhline} 
\usepackage{habbas_macros}
\newcommand{\Traces}{{\mathsf{Traces}}}

\usepackage{tikz}
\usetikzlibrary{positioning, arrows.meta, calc}
\usepackage[framemethod=tikz]{mdframed}
\usepackage{pgfplots}
\usepackage{graphicx}
\usepackage{wrapfig}
\pgfplotsset{compat=newest}
\newlength{\myline}      
\colorlet{bg}{white}
\colorlet{fg}{black}
\tikzset{
    box/.style={
        minimum size = 18pt,
        draw=fg,
        fill=bg,
        line width=\myline,
        rounded corners=2pt,
        inner sep=2pt,
        minimum width=1.8cm
    },
    wire/.style={
        line width=\myline,
        rounded corners=2pt
    },
    >={Latex[round,length=4pt,width=4pt]},
}

\newcommand{\bounds}{BOUNDS}

\begin{document}

\title{Monitoring in the Dark: Privacy-Preserving Runtime Verification of Cyber-Physical Systems}

\author{Charles Koll}
\email{kollch@oregonstate.edu}
\orcid{0000-0001-5941-250X}
\affiliation{%
  \institution{Oregon State University}
  \city{Corvallis}
  \state{Oregon}
  \country{USA}
}

\author{Preston Tan Hang}
\affiliation{%
  \institution{Hyster-Yale Materials Handling}
  \city{Fairview}
  \state{Oregon}
  \country{USA}}
\email{prestontanhang@gmail.com}

\author{Mike Rosulek}
\email{rosulekm@oregonstate.edu}
\affiliation{%
  \institution{Oregon State University}
  \city{Corvallis}
  \state{Oregon}
  \country{USA}
}

\author{Houssam Abbas}
\email{abbasho@oregonstate.edu}
\orcid{0002-8096-2618}
\affiliation{%
  \institution{Oregon State University}
  \city{Corvallis}
  \state{Oregon}
  \country{USA}
}

\renewcommand{\shortauthors}{Koll et al.}

\begin{abstract}
In distributed Cyber-Physical Systems and Internet-of-Things applications, the nodes of the system send measurements to a monitor that checks whether these measurements satisfy given formal specifications. For instance in Urban Air Mobility, a local traffic authority will be monitoring drone traffic to evaluate its flow and detect emerging problematic patterns. Certain applications require both the specification and the measurements to be \emph{private} -- i.e. known only to their owners. Examples include traffic monitoring, testing of integrated circuit designs, and medical monitoring by wearable or implanted devices. In this paper we propose a protocol that enables \emph{privacy-preserving robustness monitoring}. By following our protocol, both system (e.g. drone) and monitor (e.g. traffic authority) \emph{only} learn the robustness of the measured trace w.r.t. the specification. But the system learns nothing about the formula, and the monitor learns nothing about the signal monitored. We do this using garbled circuits, for specifications in Signal Temporal Logic interpreted over timed state sequences. We analyze the runtime and memory overhead of privacy preservation, the size of the circuits, and their practicality for three different usage scenarios: design testing, offline monitoring, and online monitoring of Cyber-Physical Systems.
\end{abstract}




\maketitle





\section{Introduction}
\label{sec:intro}
Monitoring, also known as runtime verification, is a key routine in the testing and runtime operation of Cyber-Physical Systems (CPS). 
The basic setup is one where an executing system generates an output trace $\stSig$ -- e.g. a plane engine's rotational velocity is measured, so $\stSig(t)$ is that velocity at time $t$.
A {\em monitor} receives that trace and determines whether it satisfies a formal specification of correctness.
Monitoring can be done in two different phases.
If performed at runtime (during system operation), monitoring can detect violations so the system can take remedial action.
If performed during system testing (during the design phase), monitoring is used to guide the test case generator.

In both scenarios, some applications call for privacy guarantees for both the monitored trace and the formal specification. 
In the runtime monitoring scenario, we can take the example of Urban Air Mobility, where a local traffic authority is monitoring drone traffic to evaluate its flow and detect emerging problematic patterns.
Drones' traces include positions, velocities, and even battery charges, which the drones share with the authority to enable it to safely regulate traffic.
Such flight data can reveal much about the drones' objectives and business operations. Moreover, man-in-the-middle hackers may intercept the drone-authority communication. This demonstrates a need to keep the traces private~\cite{patil24outherd}.
Conversely, because drone operators might try to game the monitor if they knew the exact specification(s) being monitored, or hackers might inject false measurements to make it look like everything is OK when it's not~\cite{sinopoli09Sec}, the specification(s) must also remain private and known only to the traffic authority.
Other examples can be given from the medical and banking sectors.
\begin{figure}[t]
  \includegraphics[width=0.5\textwidth]{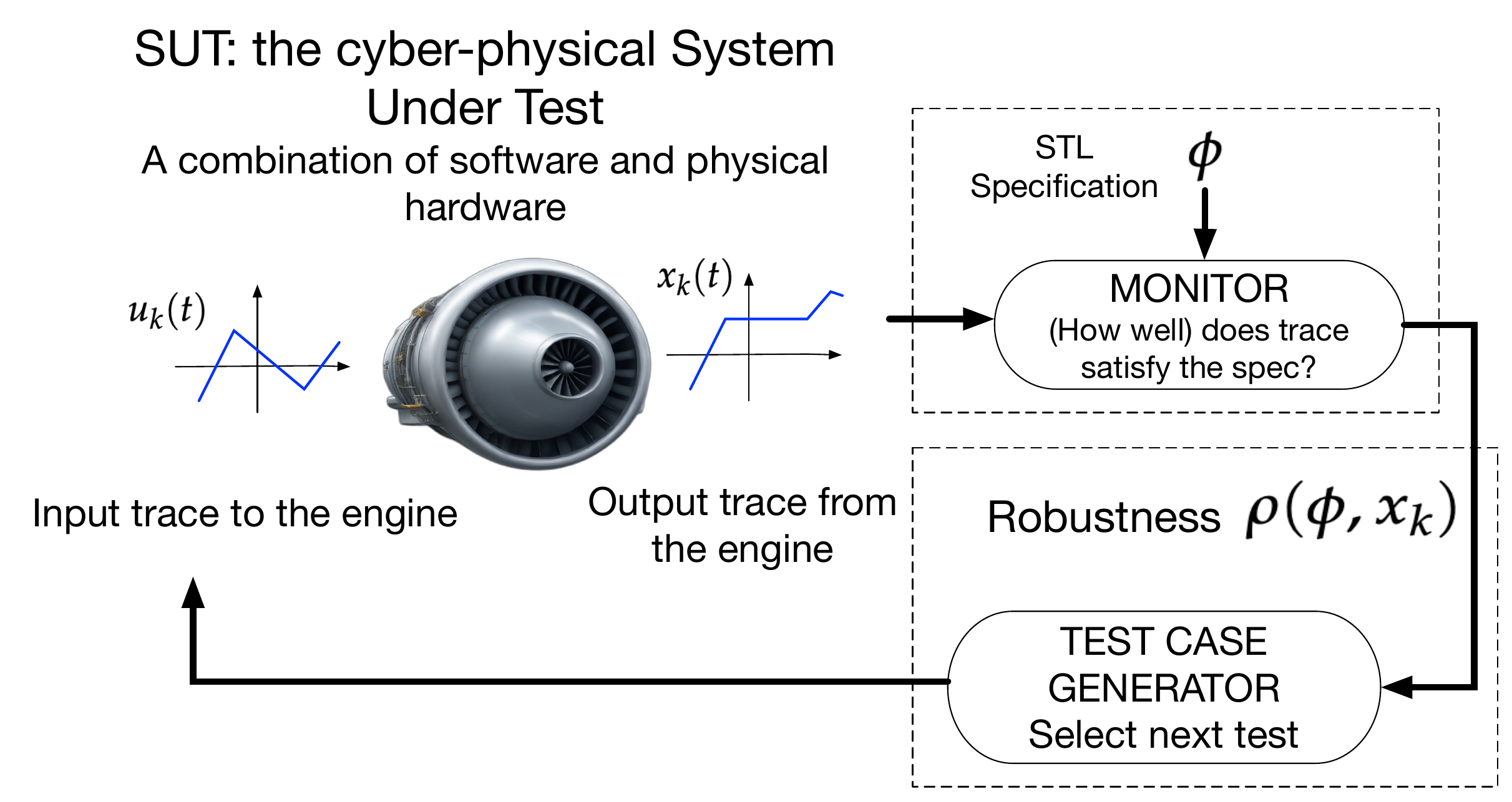}
  \caption{CPS monitoring-and-testing loop. This paper introduces a way to do the monitoring in a privacy-preserving manner.}
  \Description{CPS monitoring-and-testing loop}
  \label{fig:loop}
\end{figure}

Similar considerations apply in the testing scenario.
{\em Testing-as-a-Service} (TaaS), which is the practice of outsourcing some testing activities to an outside party, is common in industries like semi-conductor \cite{cadencesite,movellussite,rambussite}, aerospace \cite{collinstest}, fuel cell batteries \cite{avl} and software~\cite{markresearch,rainforest,testlio,councill99swtesting}.
There is significant economic incentive in building novel TaaS models: the global TaaS market in software was estimated at USD 4.54 billion in 2023~\cite{gvr24} and at USD 2.5 billions in semi-conductors~\cite{lucintel24market}.
The design company has clear incentives for keeping its design secret. The testing company possesses refined and targeted correctness specifications which it developed over years of experience with different designs. These specifications steer the test case generator towards effective tests. The testing company is thus incentivized to keep them secret. 
A key challenge of TaaS is therefore the protection of the intellectual properties (IP) of both the design and the testing companies, namely the design's output traces (and the design that produced them), and the tester's correctness specifications. 

Thus we are faced with the following problem: given two parties, Designer and Verifier, how can we enable Verifier to monitor a trace produced by Designer's system, so that both learn the outcome of the monitoring, but Verifier does not learn anything else about the monitored trace, and Designer does not learn what formula the trace is monitored against?
In this paper, we experiment with multi-party computation, and specifically garbled circuits~\cite{bellare12foundationsofgc}, to create a privacy-preserving implementation of robustness monitoring. We seek to answer the following two research questions:
\begin{enumerate}
    \item It is possible to implement private robustness monitoring in garbled circuits?
    \item If yes, is the GC implementation practical for offline testing, offline monitoring, and online monitoring?
\end{enumerate}
We show that privacy-preserving monitoring or CPS is indeed feasible with garbled circuits, and outline conditions under which it is practical.
Such a solution significantly strengthens the protection afforded by Non-Disclosure Agreements (NDAs) and physical security measures.

\paragraph{Related Work}
There are very few works on private monitoring. 
The concept of private offline \textit{LTL} monitoring was introduced in \cite{abbas19prv}, which also sketched a number of approaches to preserving the privacy of the monitored trace -- but not of the LTL formula.
Banno et al. \cite{banno22prv} define a scheme for preserving the privacy of both a {\em safety-only} LTL formula and a trace, using fully homomorphic encryption \cite{chillotti2020tfhe} and the LTL formula's representation as a DFA (Deterministic Finite Automaton).
We work with the full Signal Temporal Logic (STL) required for CPS designs, so both our signals and formulas are significantly different. 
In particular, STL formulas don't have a DFA representation, so work on oblivious DFA evaluation, like \cite{blanton10obliviousdfa} and \cite{sasakawa2014oblivious}, is not applicable here.

Some work has been done on performing (regular, non-private) monitoring with a hardware circuit \cite{jakvsic2015signal,dong2007run,reinbacher2012real}. These works generate a circuit for a given formula, whereas we design a circuit which accepts \textit{any} formula up to a given depth. Our work further uses the circuit as part of a cryptographic protocol which ensures privacy of the Designer and Verifier's inputs.

\paragraph{Contributions}
Our work provides three primary contributions:
\begin{enumerate}
    \item An approach for using garbled circuits to ensure privacy-preserved robustness monitoring of CPS;
    \item A circuit implementation of a robustness monitor, DP-TALIRO, which can handle any formula of a bounded size;
    \item An experimental evaluation of the practicality of garbled circuits-based monitoring, within a testing loop and as a stand-alone monitor. The evaluation considers monitoring runtime and memory consumption, and looks at both software implementations of GC, and potential pure hardware implementations.
\end{enumerate}

\paragraph{Organization}
The remainder of this paper is organized as follows: \autoref{sec:prelims} provides preliminaries regarding logic, robustness monitoring, and garbled circuits; \autoref{sec:problem} presents the problem statement; \autoref{sec:circuit} describes the circuit we use in the garbled circuit protocol; \autoref{sec:guarantees} states the guarantees we have with garbled circuits regarding privacy and complexity; \autoref{sec:experiments} implements the circuit, applies it to a preexisting garbled circuit package, and produces results; and \autoref{sec:conclusion} concludes.
\section{Preliminaries on Logic, Robustness Monitoring, and Garbled Circuits}
\label{sec:prelims}
\subsection{Signal Temporal Logic}
\textit{Offline monitoring} is the task of checking whether a trace satisfies a formula in temporal logic. It is called `offline' because the entire trace is available (e.g., saved in a log file).
We must define both the logic, and the traces over which its formulas are interpreted.
A trace $\stSig$ consists of a finite sequence of samples and their sampling times: $\stSig = (\stVec, \timeVec) \in [-M,M]^{N_\stSig}\times [0,T]^{N_\stSig}$, for given bounds $M$ and $T$.
This says that at \textit{timestamp} $\timeVec(i)$ the trace value is $\stVec(i)$. 
(Without loss of generality, we work with 1-dimensional signals).
The timestamps obey $0=\timeVec(0)<\timeVec(i)<\timeVec(i+1) < \timeVec(N_\stSig-1)=T$ for all $1\leq i\leq N_\stSig-3$. 
The number $N_\stSig$ of samples can differ between traces.
Bounds $M$ and $T$ are the same for all traces.
Given an interval $I \subset \nnreals$, $\timeVec^{-1}(I) = \{i \in \Ne ~|~\timeVec(i) \in I\}$.
Let $\Traces$ denote the set of all traces.

We consider specifications in Signal Temporal Logic (STL)~\cite{nickovicm07formats}. Fix a set $AP$ of atomic propositions; to every $p\in AP$ is associated a real constant $v_p$.
We write $\top$ for the Boolean constant True and $I$ for a bounded interval in $\nnreals$.
The syntax of STL is 
\[\varphi := \top~|~\stPt \geq v_p~|~\neg \formula~|~\formula \land \formula~|~\formula~\until_I\formula\]
 
Informally, atom $p$ is true of $\stSig$ at $t$ iff $\stSig(t)\geq v_p$, and $\varphi \until_I \psi$ says that $\varphi$ holds \textit{Until} some moment in $I$ at which $\psi$ holds.
Derived operators include $\formula_1 \lor \formula_2$ ($\formula_1$ or $\formula_2$), $\formula_1 \rightarrow \formula_2$ ($\formula_1$ implies $\formula_2$), $\always_I \formula$ ($\formula$ holds at every moment in $I$), and $\eventually_I \formula$ ($\formula$ holds at some moment in $I$).
Given a trace $\stSig$, a formula $\formula$, and an evaluation moment $t \in [0,T]$, one can compute whether the signal satisfies $\formula$ at $t$ (written $\stSig,t \models \formula$) or violates it (written $\stSig,t \not\models \formula$).

CPS also use a richer notion of satisfaction known as the \emph{robustness of $\stSig$ relative to $\formula$ at $i$}, written $\rob(\stSig,\varphi,i)$. Robustness is a real number $\rho$, s.t. $\rho>0$ implies that $\stSig$
 satisfies $\formula$ at $\timeVec(i)$, 
 $\rho<0$ implies that $\stSig$
 violates $\formula$ at $\timeVec(i)$.  
(The case $\rho=0$ is inconclusive for technical reasons that needn't concern us here~\cite{fainekosp09tcs}).
\emph{Thus a trace with negative robustness indicates a bug}.
The magnitude $|\rob(\trace,\varphi,i)|$ measures how well the trace satisfies or how badly it violates the formula. 
Fix an STL formula $\formula$ and trace $\stSig = (\stVec, \timeVec)$. 
For a bounded interval $I \subset \nnreals$, the notation $t+I$ is short for $\{t+s~|~ s \in I\}$.
At each integral index $i \leq N_{\stSig}-1$, the robustness is given by:
\begin{align}
\label{eq:robustness}
    \rho(\stSig,\top,i) &\defeq  \infty\\
    \rho(\stSig,p,i) &\defeq \stVec(i) - v_p,~\forall p\in AP\\
    \rho(\stSig,\neg \varphi,i ) &\defeq -\rho(\stSig,\varphi,i)\\
    \rho(\stSig,\varphi \land \psi,i) &\defeq \min(\rho(\stSig,\varphi,i),\rho(\stSig,\psi,i))\\
    \rho(\stSig,\varphi \until_I \psi,t) &\defeq \max_{j\in \timeVec^{-1}(\timeVec(i)+I)}\left(\min(\rho(\stSig,\psi,j), \min_{i\leq k < j} \rho(\stSig,\varphi,k)\right)
\end{align}

It helps the intuition to look explicitly at the Always $\always$ and Eventually $\eventually$ cases: 
\[\rho(\stSig,\always_I \varphi,i ) = \min_{j\in \timeVec^{-1}(\timeVec(i)+I)} \rho(\stSig,\varphi,j)\]
\[\rho(\stSig,\eventually_I \varphi,i ) = \max_{j\in \timeVec^{-1}(\timeVec(i)+I)} \rho(\stSig,\varphi,j)\]
If $\stSig=(\timeVec,\stVec)$, we write $\boldsymbol{\rho}(\stSig, \formula)$ for the vector of robustness values $(\rob(\stSig,\formula,0),\ldots,\rob(\stSig,\formula,N_\stSig-1))^T$. 
It is important to note that 
\[(\timeVec, \boldsymbol{\rho}(\stSig, \formula))\] 
is itself a trace.
The computation of this trace is called \textit{(offline) robustness monitoring}, and this is the computation that needs to be done in a privacy-preserving manner.

\subsection{Yao's Garbled Circuits}
Our approach to privacy-preserving monitoring is based on Yao's Garbled Circuits, or GC for short.
GC is a type of secure Multi-Party Computation (MPC), in which two or more parties collaborate to compute a common function of their secret inputs.
A tutorial introduction to GC can be found in \cite{evans2018pragmatic}, and a rigorous presentation with proofs can be found in \cite{boneh23book}. Here we present the protocol, its guarantees, and complexity.

Consider a function $F(x,y)$ that is implemented as a hardware circuit $\Cc$ with a standard set of gates (e.g. inverters, ANDs, and XORs).  Two parties, say Alice and Bob, want to compute $F(x,y)$; Alice holds secret input $x$ and Bob holds secret input $y$.
A GC is an encrypted version of $\Cc$ which allows the two parties to evaluate and learn $F(x,y)$, and learn nothing else about each other's respective inputs.
We first give the idea behind the GC construction using the simplest case of a one-gate circuit. This gate is represented by its truth table $T$, so entry $T_{xy}$ equals $F(x,y)$. If $x$ takes values in a set $X$ and $y$ in a set $Y$, then $T$ has $|X|\times |Y|$ rows. 
Call the actual inputs of Alice and Bob $x^*$ and $y^*$ respectively.

\begin{table}[ht]
    \centering
    \caption{Garbled evaluation of an AND gate with input wires $i$ and $j$ and output wire $w$. First three columns are the gate's truth table. For illustration the fourth column gives the encrypted table for the one-gate circuit: key $k_x^i$ is used for wire $i$ if it holds value $x$, and every input combination has a corresponding key combination. Fifth column gives the encrypted table in the general circuit case. The label on output wire $w$ is now itself a key, $k_x^w$, to be used for executing gates in the fan-out of this gate.}
    \begin{tabular}{|c|c|c|c|c|}
         \hline
         0&0  &0  &$Enc_{k^i_0,k^j_0}(0)$&$Enc_{k^i_0,k^j_0}(k_0^w)$ \\
         0&1  &0  &$Enc_{k^i_0, k^j_1}(0)$ &$Enc_{k^i_0, k^j_1}(k_0^w)$ \\
         1&0  &0  &$Enc_{k^i_1,k^j_0}(0)$ &$Enc_{k^i_1,k^j_0}(k_0^w)$ \\
         1&1  &1  &$Enc_{k^i_1,k^j_1}(1)$ &$Enc_{k^i_1,k^j_1}(k_1^w)$ \\ 
         \hline
    \end{tabular}
    \label{tab:gc table}
\end{table}

Alice, the \textit{garbler}, assigns randomly generated encryption keys to each possible value of $x$ and $y$, call them $k_x$ and $k_y$. She then encrypts $T$ by encrypting each element $T_{xy}$ by {\em both} $k_x$ and $k_y$, randomly permutes the rows of the table, and sends the permuted table to Bob. 
See \autoref{tab:gc table}.
Bob, the {\em evaluator}, is the party that evaluates the function.
Alice, knowing her own input $x^*$, sends $k_{x^*}$ to Bob; because the key is randomly generated, this tells Bob nothing about $x^*$.
Alice doesn't know Bob's secret input $y^*$, so they use a 1-out-of-$|Y|$ {\em oblivious transfer} (OT) protocol: this allows Bob to obtain $k_{y^*}$ without him learning anything about any other $k_y$, and without Alice learning anything about $y^*$.
Now Bob has the right keys to decrypt the desired entry $T_{x^*y^*} = F(x^*,y^*)$. But he doesn't know which row to decrypt; for this, the two parties use a {\em point-and-permute} approach, where the garbler appends to each key a pointer to the right row of the table. See \cite{boneh23book,evans2018pragmatic} for details. 
Now that he has decrypted the right entry and obtained $T_{x^*y^*}$, Bob sends that to Alice so both know the output.

We see that evaluating the one-gate circuit requires five message transmissions: OT requires three transmissions, Alice sends the encrypted table and key (one transmission), and Bob sends the final output, for a total of five. 
Alice's message has size $B|X||Y|+\kappa$, where $B$ is the number of bits to represent one entry of the encrypted table and $\kappa$ is key size (e.g., 256). 

For a general, multi-gate circuit, the garbler sends all the gates' tables and all her input keys $k_x$ to the evaluator, and they use OT to transmit the input $k_y$ keys. To avoid revealing intermediary gate outputs, the outputs themselves are encrypted (so an inner gate's table has encrypted outputs) -- see \autoref{tab:gc table}. The final outputs are sent to the garbler. 

GC is secure against both semi-honest (defined below) and malicious adversaries~\cite{evans2018pragmatic}. 

\subsection{The Real-Ideal Paradigm of Security}
We add a few words about the notion of security used in garbled circuits and Multi-Party Computation (MPC) more generally: we posit an ideal world, in which the two parties can implement the desired functionality $F$ with the help of a trusted third party $\Tc$. The two parties send their respective inputs to $\Tc$ via secure channels, and $\Tc$ computes the output $F(x,y)$ provided the input sizes are of some maximum known length (i.e. the input lengths must be less than some known upper bound, otherwise $\Tc$ rejects them). 
$\Tc$ then sends the output to the parties. 

This is an {\em ideal} situation: it is not possible for the parties to learn anything beyond what they can infer from the knowledge of the maximum input lengths, the computation $F$ being performed, their own inputs, and the output (and this much knowledge is inevitable, since the parties {\em must} know the maximum lengths, the computation being performed, their own inputs, and the final output).
A real protocol can, at best, achieve this ideal level of security. Thus we say that the real protocol is secure (or privacy-preserving) if it provides the same guarantees as the ideal world with high probability. (The guarantee is probabilistic because of the randomness introduced by encryption and possibly other steps of the protocol. The exact meaning of `high probability' is given using negligible functions and is detailed in \cite{boneh23book}).

\section{Problem Statement}
\label{sec:problem}
With the preliminaries now established, we can formally state our problem: there are two parties, Designer and Verifier. 
Designer holds a secret trace $\stSig$ and Verifier holds a secret STL formula $\formula$. Design a protocol such that the two parties learn the robustness value $\rob(\stSig,\formula,0)$, but learn nothing else about each other's inputs beyond their length.

\paragraph{System Architecture.}
A classical non-private (cleartext) CPS monitoring and testing loop is shown in \autoref{fig:loop}. This paper's focus is on the monitor.

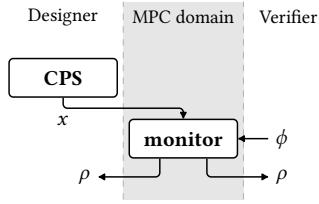
\begin{figure}
\centering
\scalebox{0.8}{
    
    \begin{minipage}{0.37\linewidth}
    \begin{tikzpicture}
        \fill[fg!10!bg] (1,0) rectangle (3,3.25);
        \draw[fg!30!bg,dashed] (1,0) -- (1,3.25);
        \draw[fg!30!bg,dashed] (3,0) -- (3,3.25);
    
        \node[box] (sut)    at (0,2)   {\bf CPS}; 
        \node[box] (mon)    at (2,1) {\bf monitor}; 
    
        \node at (0,3) {\small Designer};
        \node at (2,3) {\small MPC domain};
        \node at (3.75,3) {\small Verifier};

        \draw[wire,->] (sut) |- node[below]{$\stSig$} ($0.5*(sut.south) + 0.5*(mon.north)$)
                                -| (mon);
        \draw[wire,<-] (mon.east) -- ++(0.5,0) node[anchor=west]{$\formula$};
    
        \draw[wire,->] (mon.320) |- 
            ($(mon.south east) + (0.5,-0.3)$)
            node[anchor=west]{$\rob$};
        \draw[wire,->] (mon.220) |- 
            ($(mon.south west) + (-0.5,-0.3)$)
            node[anchor=east]{$\rob$};

        \coordinate (nw) at ($(sut.north west) + (-0.2,0.2)$);

    \end{tikzpicture}
    \end{minipage}
}
\caption{Privacy-preserving monitoring using secure multi-party computation (MPC). Designer inputs the monitored trace, Verifier enters the formula. Both learn only the robustness $\rob(\stSig,\formula,0)$.}
\Description{Privacy-preserving monitoring. Designer inputs the monitored trace, Verifier enters the formula. Both learn only the robustness.}
\label{fig:mike monitoring}
\end{figure}

\paragraph{Privacy Goals}
\label{sec:threat model}
Designer wishes to keep private the trace $\stSig$ and Verifier wishes to keep private the formalized specification $\formula$.
See \autoref{fig:mike monitoring}.

\paragraph{The Threat Model.} The threat model considers that the parties are {\em honest-but-curious}, aka {\em semi-honest}: this means that both parties follow the agreed-upon protocol, but might try to learn more than they should from the information they see.
For example if the protocol calls for the Designer to generate random bits, it will not use pre-determined bits instead -- that would be dishonest. 
However, from observing the communicated info, it might try to learn something about Verifier's formula.


\section{Circuit for Robustness Calculation}
\label{sec:circuit}
This section will describe the process of developing the Register Transfer Logic (RTL) circuit that performs robustness calculations. Verilog, a popular hardware description language, is used to blueprint the circuit's logic before it is synthesized into a gate-level circuit.
That circuit is then used as input to TinyGarble~\cite{songhori2015tinygarble}, an implementation of GC. 
TinyGarble allows sequential gates (like flops), which allows us to design a circuit with a smaller footprint than possible with purely combinational circuits.

\begin{algorithm}
    \KwIn{$\phi$ as a bus $[\phi_1, \dots, \phi_m]$, the signal $s$ as a bus $[(x_1, t_1), \dots, (x_n, t_n)]$}
    \KwOut{Return the value stored in $R[1, 1]$}
    \BlankLine

    \Proc{DP-TALIRO($\phi, s$)}{
        $b_l, b_u := [\infty, \dots, \infty], [-\infty, \dots, -\infty]$ ($m$ times each)\;
        \For{$i := n \KwTo 1$, $j := m \KwTo 1$}{
            \Switch{$\phi_j$}{
                \lCase{$\top$}{$R[i, j] := \infty$}
                \lCase{$\text{``$x > c$''} \in AP$}{$R[i, j] := x_i - c$}
                \lCase{$\text{``$x \leq c$''} \in AP$}{$R[i, j] := c - x_i$}
                \lCase{$\neg \phi_k$}{$R[i, j] := -R[i, k]$}
                \lCase{$\phi_{k_1} \land \phi_{k_2}$}{$R[i, j] := \min(R[i, k_1], R[i, k_2])$}
                \lCase{$\phi_{k_1} \lor \phi_{k_2}$}{$R[i, j] := \max(R[i, k_1], R[i, k_2])$}
                \lCase{$\phi_{k_1} \implies \phi_{k_2}$}{$R[i, j] := \max(-R[i, k_1], R[i, k_2])$}
                \lCase{$\phi_{k_1} \iff \phi_{k_2}$}{$R[i, j] := \min(\max(-R[i, k_1], R[i, k_2]), \max(R[i, k_1], -R[i, k_2]))$}
                \uCase{$\phi_{k_1} U_{I}\:\,\phi_{k_2}$}{
                    \lIf{$i = n \text{ and } 0 \in I$}{$R[i, j] := R[i, k_2]$}
                    \lElseIf{$i = n \text{ and } 0 \notin I$}{$R[i, j] := -\infty$}
                    \lElseIf{$I = [0, \infty)$}{$R[i, j] := \max(R[i, k_2], \min(R[i, k_1], R[i+1, j]))$}
                    \Else{
                        $b_l[j], b_u[j] := \mathbf{BOUNDS}(I, i, [t_1, \dots, t_n], n, b_l[j], b_u[j])$\;
                        $tmp_{min} := \min_{i \leq i' < b_l[j]} R[i', k_1]$\;
                        $R[i, j] := -\infty$\;
                        \For{$i' := b_l[j] \KwTo b_u[j]$}{
                            $R[i, j] := \max(R[i, j], \min(R[i', k_2], tmp_{min}))$\;
                            $tmp_{min} := \min(tmp_{min}, R[i', k_1])$\;
                        }
                        \lIf{$\sup I = \infty$}{$R[i, j] := \max(R[i, j], \min(R[i, k_1], R[i+1, j]))$}
                    }
                }
            }
        }
        \Return{$R[1, 1]$}
    }
    \caption{Temporal Logic Robustness Computation}
    \label{algo:dp-taliro}
\end{algorithm}

\begin{table}[ht]
    \caption{Example trace.}
    \centering
    \begin{tabular}{|c||c|c|c|c|}
        \hline
        $\timeVec$  & 0 & 5 & 7 & 10 \\
        \hline
        $\stVec$  & 3 & 11 & -2 & -3 \\
        \hline
    \end{tabular}    
    \label{tab:dp example trace}
\end{table}

\begin{table}[t]
    \caption{Constructed table for the example trace.}
    \centering
    \begin{tabular}{|c||c|c|c|c|}
        \hline
        & $\until_{[4,9)}$ & $\neg$ & $x \geq 0$ & $x \geq 10$ \\
        \hhline{|=#=|=|=|=|}
        0 ($t_1$)  & $3$ & $7$ & $3$ & $-7$ \\
        \hline
        5 ($t_2$)  & $-2$ & $-1$ & $11$ & $1$ \\
        \hline
        7 ($t_3$)  & $-\infty$ & $12$ & $-2$ & $-12$ \\
        \hline
        10 ($t_4$) & $-\infty$ & $13$ & $-3$ & $-13$ \\
        \hline
    \end{tabular}    
    \label{tab:dp example table}
\end{table}

\begin{algorithm}
    \KwIn{an interval $I$ of the form ``$[l, u)$'' where $l$ and $u$ are $\geq 0$, current index $i$, times of the signal $s$ represented as a bus $t := [t_1, \dots, t_n]$, $n$ the size of the signal, lower bound $b_l$ from the previous iteration, upper bound $b_u$ from the previous iteration}
    \KwOut{Return minimum and maximum indices relating to interval $I$}
    \BlankLine

    \Proc{BOUNDS($I, i, t, n, b_l, b_u$)}{
        \lIf{$u = \infty$}{$b_u := n$}
        \Else{
            $h_i := $ \leIf{$b_u = -\infty$}{$n$}{$b_u$}
            \For{$j := h_i \KwTo i$}{
                \If{$t_j < t_i + u$}{
                    $b_u := j$\;
                    $break$\;
                }
            }
        }
        $h_i := $ \leIf{$b_l = \infty$}{$n$}{$b_l$}
        \For{$j := h_i \KwTo i$}{
            \lIf{$t_j \geq t_i + l$}{$b_l := j$}
            \lElse{$break$}
        }
        \Return{$b_l, b_u$}
    }
    \caption{Time-stamp Bounds Computation}
    \label{algo:bounds}
\end{algorithm}

\subsection{Dynamic Program for Robustness Calculation}
We adopt the DP-TALIRO algorithm for computing robustness \cite{yang2013dynamic}. DP-TALIRO was designed for MTL robustness monitoring, but is trivially modified for STL. The algorithm contains two components: the primary robustness calculation, and a subroutine \bounds{} that selects which elements of a trace fall within the bounds of a temporal interval. The algorithm for the primary DP-TALIRO routine is provided in \autoref{algo:dp-taliro} and the algorithm for the \bounds{} subroutine is provided in \autoref{algo:bounds}. Both have been re-factored for clarity. Correctness of DP-TALIRO is shown in \cite{yang2013dynamic}.

DP-TALIRO is a Dynamic Program (DP), which is illustrated in the example below (Tables \ref{tab:dp example trace} and \ref{tab:dp example table}).
The primary robustness calculation fills a table of robustness values which has one column per sub-formula, and one row per time-step in the trace. 
The column headers are sub-formulas, appearing in a breadth-first traversal order of the formula's syntax tree (thus $\formula$ heads the first column and the atoms head the last columns).
Time increases along the rows; see \autoref{tab:dp example table}.
Thus the table has size $N_\stSig \cdot M$, where $N_\stSig$ is the trace length and $M$ is the number of nodes in the formula's syntax tree. 
The table is filled starting from the bottom right and going up to the top left. Table entry $[i,\psi]$ contains robustness value $\rob(\stSig,\psi,i)$, where $\psi$ is the sub-formula that heads that column. 

There are five different components in the STL syntax, each processed in a unique way, as per the robust semantics in \autoref{sec:prelims}: True ($\top$) results in $\infty$, Atomic Propositions return $\stVec(i)-v_p$, Negation ($\neg \formula$) returns the negation of the robustness of its child formula $\phi$, Conjunctions ($\formula_1 \land \formula_2$) return the minimum robustness between their two child formulas $\formula_1$ and $\formula_2$, and Until ($\formula_1~\until_I \formula_2$) requires more computation which is illustrated below, involving the subroutine \bounds. After completion of the robustness table, the final value located at entry $[0,0]$, which equals $\rob(\stSig,\formula,0),$ is returned.

The subroutine \bounds{} takes two arguments, an index $i$ into the trace and an interval $I \subset \nnreals$, and returns the minimum and maximum values in $\timeVec^{-1}(\timeVec(i)+I)$. It does this by simply iterating through the trace starting at $i$. Once found, these two indices are passed to the primary robustness calculation to be used for an Until.

\paragraph{Example}

Now we provide an example calculation of the dynamic programming algorithm. Suppose we have a trace as shown in \autoref{tab:dp example trace} and a formula of \autoref{eq:dp example formula}:
\begin{equation}\label{eq:dp example formula}
    \phi: (x \geq 0)~\until_{[4,9)} \neg(x \geq 10)
\end{equation}

We can construct a table with the headers of \autoref{tab:dp example table} and fill in the table by following the DP-TALIRO algorithm. Our final robustness result of the trace $x$ (\autoref{tab:dp example trace}) relative to formula $\phi$ (\autoref{eq:dp example formula}) is the top left cell of the table.

We begin filling in the table by starting in the bottom right cell, moving left across the bottom row. When evaluating, we consider only the trace suffix from timestep $t=10$ onwards, which in this case is merely the point $\stSig(10)= -3$. For the rightmost cell, we compute the robustness $-3 - 10 = -13$ (recorded in \autoref{tab:dp example table}). The next cell to the left is computed similarly: $-3 - 0 = -3$. The cell after that, the third from the right, is the first which requires referring to one of our previously-computed cells. In the sub-formula the negation is applied to $x \geq 10$, meaning that we need to look at the cell in the current row which corresponds to $x \geq 10$, the first one we computed. Then our calculation is $-(-13) = 13$. Finally in this row we compute the Until: \bounds{} determines that no samples after $t=10$ fall within the the $[4,9)$ interval (since $t=10$ is the last sample).  
The result of this computation is $-\infty$.

We now move up to the penultimate row and perform similar operations. The computations of the first three cells from right to left are $-2 - 10 = -12$, $-2 - 0 = -2$, and $-(-12) = 12$, respectively. \bounds{} determines again that no samples fall within $7+[4,9)$ and again the result is $-\infty$. 
When we reach the second row, on the other hand, \bounds{} determines that sample $t=10$ falls within $5+[4,9)$, so the relevant table cells contain the values 11 and -2 (from the $x\geq 0$ column) and 13 (from the $\neg$ column). The Until returns the final value -2. For details on the Until computation, see \cite{yang2013dynamic}.

The algorithm continues to work upwards through the rows until the final cell, the top left of the table, is computed. This cell is the result of the entire DP-TALIRO algorithm, being the robustness of the trace relative to the formula.

\subsection{RTL Design}
DP-TALIRO is a dynamic program without recursive calls, iteratively computing values until a final result. It has been implemented by various authors in C, Python, and Rust. 
For this work, we implemented the algorithm in Verilog, a high-level hardware description language with C-like constructs. Verilog is the dominant language in the semi-conductor industry for hardware design and verification.

Our implementation, following common best practices, uses a controller+datapath design separation. The controller is a finite state machine used for control logic, like the ``switch'' statement in \autoref{algo:dp-taliro} line 4. The datapath handles the calculations of the algorithm, such as the negation on line 8 or min calculation on line 9.
In our implementation the datapath is the component which takes the circuit input -- both trace and formula -- and returns the circuit output, the robustness result.
To be clear, we do not produce a different Verilog implementation for different formulas - there is no `compilation' step. 
This is an implementation of DP-TALIRO in Verilog, much like others have implemented it in Python or C.

\begin{figure*}[ht]
  \includegraphics[width=\textwidth]{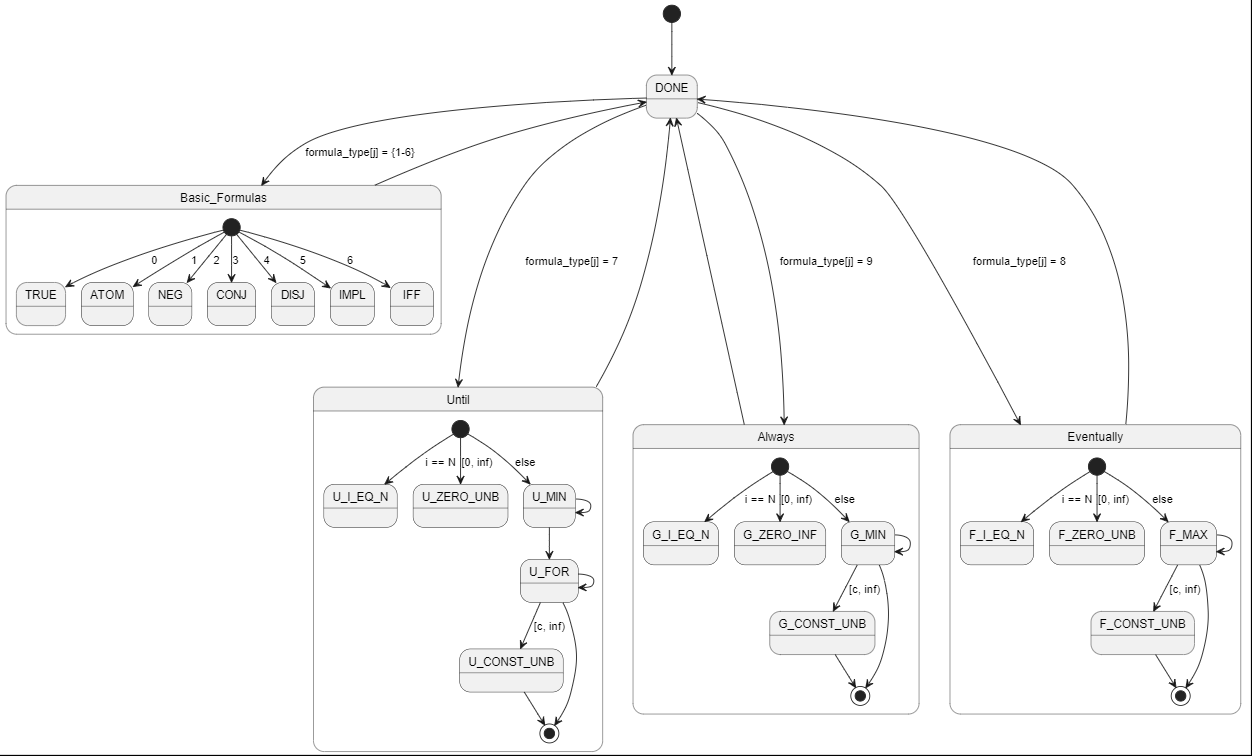}
  \caption{Finite state machine of the controller. Temporal blocks are labeled with conditions on their intervals (e.g. $U\_ZERO\_UNB$ corresponds to $\until_{[0,\infty)}$ and $F\_I\_EQ\_N$ for Eventually when the current table index equals signal size $N$).}
  \Description{Finite state machine diagram of the Control}
  \label{fig:fsm}
\end{figure*}

The trace input is represented as pairs $(t,x)$ of 32-bit signed integers, with $t$ being the timestamp and $x$ the signal value. 
The formula input has two components -- an array of labels for operators and atomic propositions, and the corresponding parameters. 
The array of labels encodes the position of the operators and atoms in the parse tree of the formula.
The parameter associated to a temporal operator is simply the temporal interval $[a,b]$, represented by a pair of 32-bit signed integers. 
For an atomic proposition the parameter is the value $v_p$.

The controller is implemented as a finite state machine (FSM) with 21 states, shown in \autoref{fig:fsm}. The FSM is 1-hot encoded for speed.
First, the input formula (and its sub-formulas) are classified as being one of four formula types: Basic Calculations ($\top$, atomic propositions, and Boolean operators) are one type, as they can all be computed in one clock cycle, allowing the control to command a single state transition to the datapath.
The three other types correspond to temporal operators (Until, Always, and Eventually), which require more clock cycles. 
Each temporal operator contains three branches, corresponding to lines 14-15, 16, and 17 in the algorithm.
The first branch is taken if this the start of the computation for this operator (indicated by the current calculation index $i$ equaling $N$, lines 14-15), the second if the temporal operator's interval is $[0,\infty)$, and the last if the temporal operator's interval is any other possibility $[a,b)$. Operations in the third branch require looping calculations. 
Since a single loop iteration occurs in one clock cycle, a state that contains a transition to itself is required to achieve multiple loops.

The datapath module performs all calculations through sequential logic, as flip-flops are needed for the robustness values.
The datapath is designed with two main blocks: one that handles data processing and one that calculates the robustness for the current table indices.

\begin{figure*}[ht]
    \includegraphics[width=0.7\linewidth]{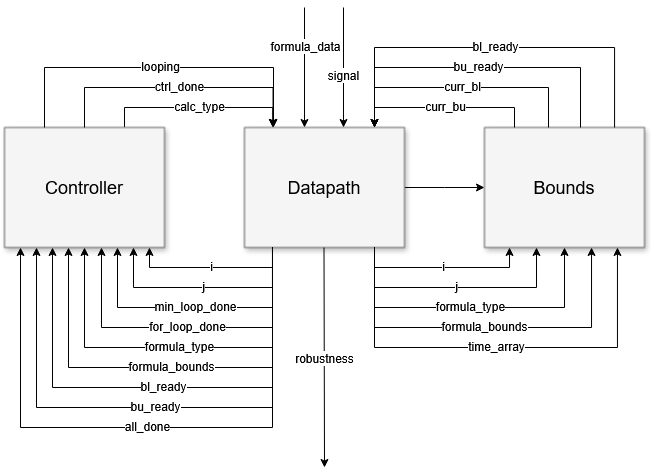}
    \caption{Movement of data between the Datapath, Controller, and Bounds modules. The Datapath handles input and output for the entire algorithm, taking in formula data and the signal and returning the robustness. Variables being passed to and from the Controller inform the Datapath on which calculations to perform next, and variables from the Bounds module provide information on which parts of the signal to consider in its calculations, given a subformula and starting point of the signal. Clock and reset variables are not included in this figure.}
    \Description{Flowchart with three cells: Controller, Datapath, and Bounds. Both Controller and Bounds have labeled arrows from and to the Datapath. There are two labeled arrows starting from outside the figure, pointing at the Datapath: formula_data and signal. There is one labeled arrow originating from the Datapath and pointing outside the figure: robustness.}
    \label{fig:circuit overview}
\end{figure*}

An additional {\sffamily bounds(...)} module implements the \bounds{} routine to determine the upper and lower limits of temporal intervals that appear in the DP table (e.g. $\timeVec^{-1}(\timeVec(i)+I)$ in \autoref{eq:robustness}). It completes entirely before the rest of the datapath begins processing the (sub-)formula, granting immediate access to interval bounds during robustness calculations.

Communication between the controller and datapath occurs through numerous 1-bit flags, indicating things like progress through a computation, end of a computation and reset. \autoref{fig:circuit overview} provides more information on the data passed between the modules.

The Verilog description is synthesized into a gate-level netlist by Synopsys' Design Compiler (DC) tool. Circuits sizes are given in the Experiments section.


\section{Theoretical Guarantees}
\label{sec:guarantees}
\subsection{Privacy Guarantees}
The privacy guarantees are automatically obtained through the use of garbled circuits.
The GC protocol has been shown to be secure \cite{bellare12foundationsofgc}, i.e. with high probability, each party does not learn anything beyond their own input, the output of the circuit, the lengths of the inputs, and the circuit itself.

Given this notion of security, the lengths of inputs and outputs are known by both parties. This includes both the length of the trace and the depth of the formula. Note however, that a party with a shorter input can pad it to fit the circuit -- in that case, the other party does not learn the true length of the input. As an example, consider a case where both parties have agreed to use a circuit which expects a formula with depth 4, but the Verifier has a formula of depth 3. Then when providing the shorter formula to the circuit, the Verifier can pad the remaining input bits with 1; these bits do not impact the robustness calculation, regardless of the trace.

\subsection{Complexity}
\subsubsection{Circuit size}
The size of the circuit depends on the number of AND gates required to compute a single clock cycle, since only AND gates require an encryption lookup table (e.g. \autoref{tab:gc table}). The only other gates in GCs are inverters and so-called `free XORs'.
This does not mean that the number of {\em encryptions} is equal to the number of AND gates: TinyGarble encrypts the gates {\em for each clock cycle in which they are evaluated}. Thus by supporting sequential elements, TinyGarble enables a relatively small circuit size $C$ but the total effective size of the circuit for the GC protocol relies on the number of clock cycles $cc$ needed to perform the computation: $O(C \cdot cc)$.

\subsubsection{Number and size of message transmissions}
GC has a constant number of message transmissions between the two parties. The total size of these transmissions is $O(C \cdot cc \cdot \kappa)$ bits, where $\kappa$ is the security parameter (commonly 256 bits). This is because all gates in the circuit must be encrypted by the garbler and sent to the evaluator, and TinyGarble does this for each clock cycle. The encryption size of a gate is only reliant on the security parameter $\kappa$ and is not impacted by the number of gates.

\subsubsection{Memory consumption (RAM)/Space complexity}\label{sub:space complexity}
Space complexity is linear in the size of the circuit: $O(C \cdot \kappa)$. In classical GC, assuming a combinational circuit, all encrypted gates are sent from the garbler to the evaluator in a single message.
In a sequential circuit, only the encrypted values of the gates from the previous clock cycle are needed. This means that the space complexity is $O(C \cdot \kappa)$ and not affected by the number of clock cycles $cc$.

\section{Empirical Evaluation}
\label{sec:experiments}

Recall our two research questions:
\begin{enumerate}
    \item It is possible to implement private robustness monitoring in garbled circuits?
    \item If yes, is the GC implementation practical for design testing, offline monitoring, and online monitoring?
\end{enumerate}

The previous sections described our GC implementation of DP-TALIRO, thus answering the first question affirmatively.
The theoretical guarantees of GC establish the security of this implementation.
In this section, we relay the results of our implementation in terms of runtime, peak memory usage, and circuit size. These allow us to determine where this GC implementation is practical. 

We synthesized 16 circuits with various input lengths: formulas of depths 3 and 4, and traces of lengths 10, 20, 50, 100, 200, 300, 400, and 500, using every pairing of formula and trace. 
For the trace data we randomly generated pairs $(\timeVec(i), \stVec(i))$, where the differences $\timeVec(i) - \timeVec(i-1)$ and $\stVec(i) - \stVec(i-1)$ are uniformly distributed. 
The input to the circuit is represented as integers, as though every signal value and timestamp were multiplied by some large multiple of 10 to make it an integer. E.g. instead of 1.36, the input is 13600 (scale by $10^5$).
This does not change the computation as long as everything is scaled: signal values and atomic proposition thresholds, timestamps and temporal interval endpoints. 
For a fine resolution of robustness values we used a wide range for the differences: $\timeVec(i) - \timeVec(i-1)$ is uniformly distributed over $[1, 2\text{ million} - 1]$ and $\stVec(i) - \stVec(i-1)$ is uniformly distributed over $[-2\text{ million}, 2\text{ million}]$. 
This produces greater variability in the data, leading to a larger number of possible robustness values.

For formula generation, we randomly selected a formula template, temporal intervals for the template, and atomic predicates for the template. The template was selected out of four options: for depth 3-formulas, we selected templates from:
\begin{itemize}
    \item $\always_I \eventually_J p$
    \item $(\neg p) \until_I (\always_J q)$
    \item $(\always_I p) \rightarrow (\eventually_J q)$
    \item $\always_I (p \land q)$
\end{itemize}
And for depth 4-formulas:
\begin{itemize}
    \item $(\always_I (p \lor q)) \until_J r$
    \item $(\always_I \eventually_J p) \rightarrow (\always_K \eventually_L q)$
    \item $\eventually_I (p \until_J (q \rightarrow r))$
    \item $\eventually_I \always_J \neg p$
\end{itemize}
Once the template was selected, temporal intervals $I, J, K, L$ were randomly set to the following:
\begin{itemize}
    \item $[0, \infty)$
    \item $[1\text{ million}, 2\text{ million})$
    \item $[0, 3\text{ million})$
    \item $[5\text{ million}, 20\text{ million})$
    \item $[600\text{ million}, 700\text{ million})$
    \item $[30\text{ million}, 31\text{ million})$
\end{itemize}
Atomic predicates were similarly chosen, replacing $p, q, r$:
\begin{itemize}
    \item $x > 5\text{ million}$
    \item $x > 0$
    \item $x > -3\text{ million}$
\end{itemize}
The large interval and predicate values were chosen to have a similar scaling as the timestamps and values in the trace.

We described our RTL circuit using the Verilog hardware description language, and synthesized it using Synopsys' Design Compiler (DC) Version U-2022.12-SP7. Circuits were synthesized on a x86-64 machine with Intel Xeon CPU X5650@2.66 GHz and 94 GB RAM.
For the garbled circuit execution, we used TinyGarble \cite{songhori2015tinygarble}, an implementation for GC which supports sequential circuits.
RTL circuits were simulated on a x86-64 machine with AMD EPYC 7313 16-Core Processor and 50 GB RAM. Experiments using TinyGarble were run on this same machine.

Here we remind the reader that {\em the formula is an input to our circuit}: thus the circuit can handle any formula up to a given depth. So the results here are not comparable to prior work on hardware-based monitoring, which synthesizes a circuit for a \textit{given} formula: such a formula-specific circuit does not have to parse the structure of the formula, nor deal with any combination of temporal intervals on the operators. By contrast our circuit has to deal with such variability to preserve the secrecy of the formula.
We also note that our circuit is significantly bigger than benchmarks that appear in the GC literature, because it implements a notably more complex functionality. 
For instance our circuit, for traces of length 200 and formula depth 4, has over 2 million wires, while the largest benchmark in the recent \cite{mo2023haac}, which implements Bubble Sort, has just under 12,500 wires.

\begin{figure}[ht]
    \centering
    \includegraphics[width=0.7\linewidth]{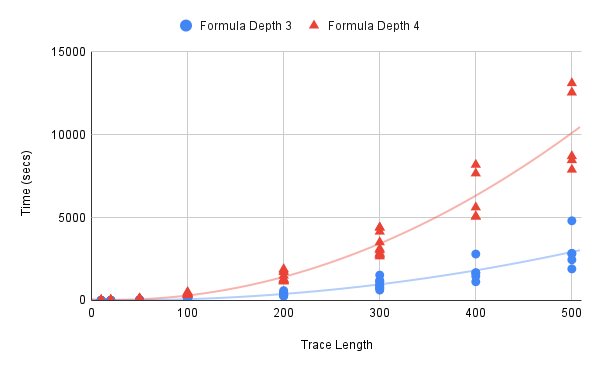}
    \caption{Monitoring runtime vs trace length. For each length, we tried multiple formulas of depths 3 and 4. The curves show the best fit.}
    \Description{Monitoring runtime vs trace length.}
    \label{fig:monitor runtime vs trace length}
\end{figure}

\begin{figure}[ht]
    \centering
    \includegraphics[width=0.7\linewidth]{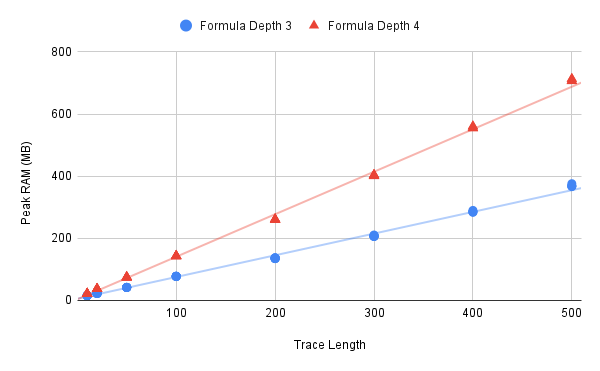}
    \caption{Monitoring peak memory consumption vs trace length. For each length, we tried multiple formula of depths 3 and 4 (many points in the graph overlap). The curves show the best fit.}
    \Description{Monitoring peak memory consumption vs trace length.}
    \label{fig:memory vs trace length}
\end{figure}

We ran ten different formulas for each trace length up to 300 and five different formulas for trace lengths 400 and 500 -- this involved running TinyGarble for a given input size, passing in the trace and formula data and measuring runtime (\autoref{fig:monitor runtime vs trace length}) and peak memory consumption (\autoref{fig:memory vs trace length}). We use these two metrics to determine feasibility of GC for the scenarios of testing, offline monitoring, and online monitoring.

\textbf{Runtime in software.}
\autoref{fig:monitor runtime vs trace length} shows the TinyGarble runtime against input trace length.
It indicates that the time to compute robustness scales exponentially with the size of the inputs, both trace and formula. Most measurements for a given trace length are fairly clustered together, with higher variance as one increases trace length. This is likely due to the number of clock cycles required for the circuit to produce a result: a higher number of clock cycles means that the circuit is effectively larger for GC, leading to differences in the amount of work done across different formulas even when these formulas have the same depth. 
The runtimes suggest that when executed in software, this approach to private runtime monitoring is practical only for smaller trace lengths and/or shallower formulas (e.g., trace length 10 and formula depth 3: runtime of approximately 1 second). 
For falsification, this will depend on how long it takes to simulate/execute the system-under-test: some systems take hours (e.g. SPICE circuits) and others a few seconds, so the overhead of private monitoring has to be measured against that.

\begin{figure}[ht]
    \centering
    \includegraphics[width=0.7\linewidth]{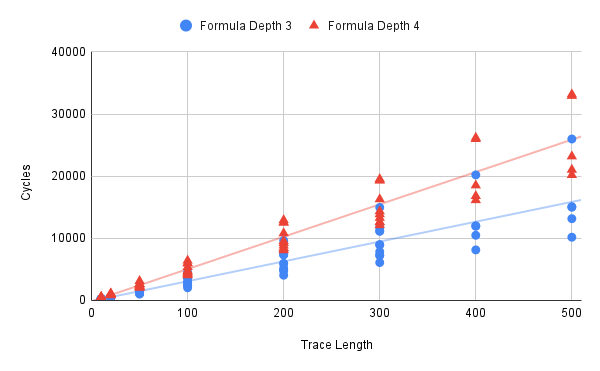}
    \caption{Hardware cycles vs trace length. For each length, we tried multiple formula of depths 3 and 4. The curves show the best fit.}
    \Description{Monitoring runtime vs trace length.}
    \label{fig:cycles vs trace length}
\end{figure}

\textbf{Hardware clock cycles.}
The above runtime results have been simulated through an OS. Work has been done on integrating GC into hardware (e.g. \cite{mo2023haac}). In such a setup, runtime scaling could potentially be sub-exponential.
\autoref{fig:cycles vs trace length} shows the number of hardware clock cycles required for the computation to complete, for each input trace length. The number of clock cycles is the most relevant metric for actual runtime once the hardware is manufactured, and must be complemented with the runtime of the GC hardware. 
For example, 40,000 cycles running on a 2.1GHz processor take 19ms.
In \autoref{fig:cycles vs trace length}, the differences in clock cycles for a given trace length are due to the specific formula being used: some formulas require additional clock cycles to compute robustness. To avoid leaking information about the provided formula through the number of clock cycles, an industrial implementation would run for a fixed number of cycles regardless of the formula. Here we report the minimum as a better metric of actual runtime once the circuit is manufactured.

\textbf{Peak memory.}
\autoref{fig:memory vs trace length} shows that the peak RAM usage of TinyGarble is linear as trace length increases and that variance for a given input size is very small (each ``point'' in this figure is 5-10 points stacked on top of each other). These results indicate that the number of clock cycles does not influence the space needed to compute the circuit -- computing at each new clock cycle only relies on the values of the gates at the previous cycle. The linear aspect of the RAM usage matches theoretical complexity (\autoref{sub:space complexity}). The peak RAM necessary for our largest-input tests suggests that embedded devices with 1 GB RAM could handle inputs of a trace of length 500 and a formula of depth 4. Shorter inputs (likely wanted on embedded devices due to lengthy runtimes) would naturally require less RAM.

\begin{figure}[ht]
    \centering
    \includegraphics[width=0.7\linewidth]{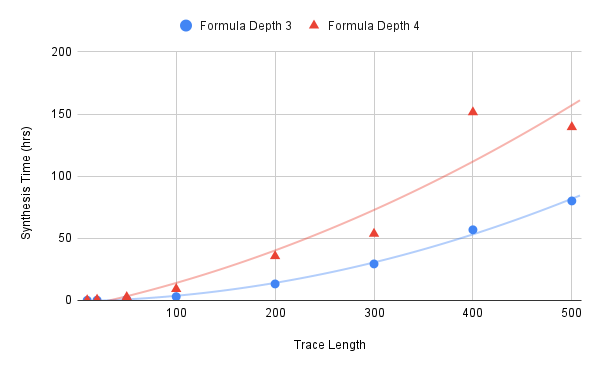}
    \caption{Synthesis time in hours for each trace length, across formulas of depths 3 and 4.}
    \Description{Synthesis time in hours for each trace length, across formulas of depth 3 and 4.}
    \label{fig:synth time vs trace length}
\end{figure}

\begin{figure}[ht]
    \centering
    \includegraphics[width=0.7\linewidth]{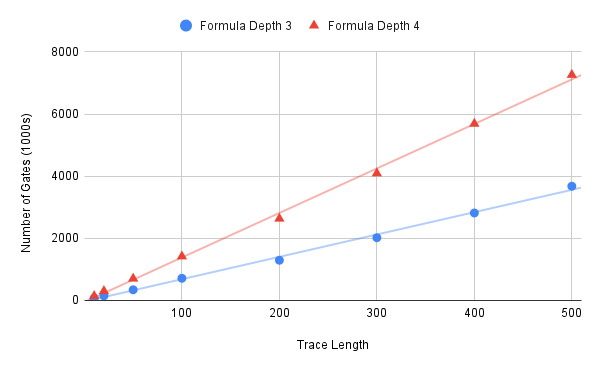}
    \caption{Size of the synthesized circuit, measured by number of gates (in 1000s). Compared across formulas of depth 3 and 4.}
    \Description{Size of the synthesized circuit, measured by number of gates (in 1000s). Compared across formulas of depth 3 and 4.}
    \label{fig:synth size vs trace length}
\end{figure}

\textbf{Circuit size.}
A gate-level netlist is synthesized once from the Verilog, and then it can be garbled repeatedly. \autoref{fig:synth time vs trace length} shows the time it takes to synthesize the gate-level netlist and \autoref{fig:synth size vs trace length} shows the netlist's size. The synthesis time scaled exponentially with respect to the size of the inputs. Note that the synthesis time is measured in hours -- for the trace lengths 400 and 500 and formula depth 4, it took nearly a week for the circuit to be synthesized. This in itself is not a significant concern since synthesis only has to occur once, but for larger input sizes this may become intractable. Note also that synthesis occurs offline, prior to any testing or monitoring; this means that synthesis can be done on more powerful computers, at which point the synthesized circuit can be loaded onto the CPS.

The size of the synthesized circuit scales linearly in the trace length. This is expected since the length of the trace determines the number of rows in the robustness table (see \autoref{tab:dp example table}). Each new row adds a fixed number of gates to the circuit.

\paragraph{Discussion.}
The runtimes suggest that GC-based monitoring can be practical in TaaS (i.e., as part of a testing loop), in most scenarios if trace length is below 200. Above that, it depends on the time it takes to simulate the design: if generating a trace takes a significant amount of time (say, a few hours, as can happen for transistor-level simulations, for example) then monitoring time would be comparatively low and an acceptable cost.
Offline monitoring, likewise, can use our approach when one long trace, or many short traces, are monitored, and the system conducting the monitoring has up to 1 GB of RAM.
Online monitoring would find GC impractical for any input sizes beyond short traces (up to 50 time steps approximately) and simple formulas. Larger input sizes would be limited by monitoring runtime, since RAM constraints appear to scale linearly.

If GC is fully implemented in hardware, its applicability scales significantly.
Synthesis time for a given circuit is significant regardless of the input sizes, but as noted above, each circuit only needs to be synthesized once, and can be used any number of times, since its structure is independent of the monitored formula and trace.

\section{Conclusion}
\label{sec:conclusion}
We presented the first privacy-preserving robustness monitor for STL formulas. Our monitor works over timed state sequences, and used garbled circuits as the underlying multi-party protocol for preserving privacy.
We characterized empirically the size of the gate-level circuit as a function of trace length and formula depth. In particular, synthesizing a circuit for a trace length greater than 500 data points is prohibitive, calling for a more sophisticated approach to robustness computation, perhaps conservatively sacrificing some accuracy to get a smaller circuit.
Execution time in software scales exponentially with the trace length, but hardware-based implementations of garbled circuits offer a possibility that practical implementations can be developed for runtime monitoring and falsification, particularly in system testing.



\bibliographystyle{ACM-Reference-Format}
\bibliography{references}

\end{document}